\documentclass[11pt,a4paper,english,superscriptaddress,aps,nofootinbib]{revtex4}

\usepackage{color}
\usepackage{amsmath}
\usepackage{amssymb,amsfonts,latexsym,cancel} 
\usepackage{graphicx}
\usepackage{hhline}
\usepackage{mwe}
\usepackage{amsbsy}
\usepackage{textcomp}
\usepackage{commath}
\usepackage{subcaption}
\usepackage{float} 
\usepackage{mathrsfs}
\usepackage{bm}
\usepackage{array} 
\usepackage{ragged2e}

\DeclareMathOperator{\sech}{sech}

\usepackage[english]{babel}
\newcommand{\bea}{\begin{eqnarray}}
\newcommand{\eea}{\end{eqnarray}}

\newcommand{\pa}{\partial}

\newcommand{\be}{\begin{equation}}
\newcommand{\ee}{\end{equation}}

\numberwithin{equation}{section}

\makeatletter

\begin{document}

\title{Sphaleron decay on circles}
\date{\today}

\author{S. Navarro-Obregón}
\email{sergio.navarro.obregon@uva.es}
\affiliation{Departamento de Física Teórica, Atómica y Óptica and Laboratory for Disruptive Interdisciplinary Science (LaDIS), Universidad de Valladolid, 47011 Valladolid, Spain}

\author{J.M. Queiruga}
\email{xose.queiruga@usal.es}
\affiliation{Departmento de Matemática Aplicada, Universidad de Salamanca,
Casas del Parque 2, 37008 - Salamanca, Spain}
\affiliation{Instituto Universitario de Física Fundamental y Matemáticas, Universidad de Salamanca, Plaza de la Merced 1, 37008 - Salamanca, Spain}

\begin{abstract}
We discuss various sphaleron-like solutions on $\mathbb{S}^1$. These solutions are static, but unstable. We explore possible stabilization mechanisms based on the excitation of internal modes. Additionally, we observe that, on time scales comparable to the size of the circle, the collapse of large sphalerons mimics the kink-antikink scattering on the real line.
\end{abstract}

\maketitle


\section{Introduction}
\label{intro}

Sphalerons are static, unstable solutions of certain nonlinear field theories that arise when the configuration space admits noncontractible loops \cite{Manton-WS, Forgacs}. They may exist in theories without topological sectors, therefore, in some sense, sphalerons are more general objects than topological solitons. In addition, they play a relevant role in physics. For instance, the electroweak sphaleron can be interpreted as a saddle point in a family of static field configurations that interpolate between two vacua with distinct winding numbers \cite{Manton-WS, Manton-Klinkhamer}. In contrast to quantum vacuum tunneling processes mediated by instantons, which are exponentially suppressed, the sphaleron can mediate processes that violate baryon and lepton numbers, provided that sufficient energy (greater than the sphaleron energy) is available \cite{Manton-Klinkhamer, Rubakov-current,Rubakov-Universe, Bochkarev-AH, Carson-AH}.

The instability of the sphaleron can be attributed to the presence of negative modes in the linear spectrum of perturbations \cite{Brihaye-modes,Carson-modes}. Specifically, the existence of $n$-negative modes corresponds to the presence of 
$n$-unstable directions \cite{Yaffe-instabilities,Akiba-instabilities}. The half-life of the sphaleron is directly related to the time scale determined by these negative modes. In certain cases, the presence of positive modes in the spectrum can significantly influence the sphaleron's half-life \cite{Sergio-phi6}. For instance, we will show that if some of these modes are excited, they can generate an effective potential that temporarily stabilizes the sphaleron.

Probably, the simplest one-dimensional models containing sphalerons are real field theories with false vacua \cite{Bazeia-false,Bazeia-false2,Alberto-false,Manton-phi3}. In such models, the sphaleron connects the false vacuum to itself and its instability can be traced back to the presence of the false vacuum.   
Maybe, a more interesting situation in $d=1+1$ arises in models whose base space manifold is compact, e.g., a circle, since quantum field theories on a finite volume are intimately related to quantum field theories at finite temperature. The easiest examples on the circle are the $\phi^4$ model \cite{Manton-circle} and the sine-Gordon model \cite{Jiu-modes}. The existence of sphalerons in these models can be understood as follows: assume that a kink-antikink ($K\bar{K}$) pair is created from one of the vacua. This configuration may travel around the circle in opposite directions and annihilate, leaving behind the other vacuum. There is an intermediate configuration where the $K\bar{K}$ pair is located at two opposite points on the circle. At this point, the attraction through one of the vacua exactly compensates for the other and the solution becomes static. This is precisely the sphaleron solution. Actually, this is the simplest one, since it is easy to show that the $n$-pairs of $K\bar{K}$s equidistributed along the circle will also generate static solutions. Under this interpretation, the sphaleron can be realized as a static solution connecting two distinct vacua. 

In one-dimensional models, sphalerons have generically an unstable mode. Excitations of this mode cause the sphaleron to decay to one of the vacua. In theories with reflection symmetry, where kinks and antikinks are related by a spatial inversion, such as $\phi^4$ or sine-Gordon, the decays to any vacuum are equivalent in the sense that they are related by the symmetry. A particularly interesting scenario arises in models that lack reflection symmetry, such as the $\phi^6$ theory. In this case, we will show that the decay of the sphaleron to distinct vacua produces entirely different evolution patterns.

In situations where the size of the base space manifold is large enough, one can interpret the sphaleron as a $K\bar{K}$ configuration on the real line. As a consequence, one should expect that, at least in time scales comparable to the size of the base space, the evolution of the sphaleron decay should share some similarities with the $K\bar{K}$ scattering patterns. In this work, we will show that this is exactly the case. However, since the base space is a compact manifold, at very large times the evolution should be rather chaotic, probably given rise to a thermalized state. The exception to this behavior is given by integrable theories. We will show, for instance, that the sine-Gordon sphaleron evolves into an exact periodic solution right after the decay.  

Finally, we will also consider in detail the influence of positive internal modes on the subsequent evolution of the sphaleron \cite{Sergio-phi6}. The excitation of the positive internal modes during the sphaleron decay may generate an effective force that stops the collapse, modifying the lifetime of the sphaleron. This could be a generic mechanism that modifies the sphaleron lifetime. 

It is the purpose of this paper to describe the evolution of some one-dimensional sphalerons on the circle. In Sec. \ref{sec:General} we explain some generalities of $1+1$ dimensional scalar field theories on the circle. In Sec. \ref{sec:phi4-SG-phi6} we review the $\phi^4$, $\phi^6$ and sine-Gordon sphaleron and in Section \ref{sec:decay} we analyze their decays. In Sec. \ref{sec:stab} we study the influence of the positive internal modes on the lifetime of the sphaleron. Finally, Section \ref{sec:conclusions} is devoted to our conclusions.


\section{$1+1$ dimensional scalar field theories on the circle} \label{sec:General}

Let us start considering a general $1+1$ dimensional scalar field model on the circle with the following Lagrangian density
\be \label{Lag_sph}
\mathcal{L}=\frac{1}{2}\pa_\mu\phi\pa^\mu\phi-U(\phi).
\ee
The field $\phi$ is a map that satisfies
\be\label{eq_period}
\phi: \mathbb{S}^1\rightarrow \mathbb{R},\quad \phi(t,x)=\phi(t,x + L),
\ee
where $L$ is the length of the circle. We assume that $U(\phi)$ is a positive semidefinite potential with $n$ vacua, denoted by $\phi_i$ and, for simplicity, we also assume that the vacua are zeros of the potential. 

The equation of motion associated to (\ref{Lag_sph}) is
\be\label{second_eq}
\square \phi +U'(\phi)=0,
\ee
which can be trivially integrated once
\be\label{first_eq}
\frac{1}{2}\phi'(x)^2-U(\phi)=-\frac{1}{2}C^2,
\ee
where $C$ is a real constant. Let us call $\phi_i^{(m)}$ the position of the greatest maximum of the potential $U$. Then, it is easy to see from (\ref{first_eq}) that the periodicity condition can only be satisfied for $0\leq C^2 \leq 2 \,U(\phi_i^{(m)})$. The first order equation (\ref{first_eq}) has two types of trivial solutions satisfying (\ref{second_eq}) for all lengths $L$. The first type of solutions are the vacua of the theory
\be
\phi(x)=\phi_i, \,\, \text{with $C=0$},
\ee
where $E(\phi_i)=0$. The second type are the unstable constant solutions
\be
\phi(x)=\phi_i^{(m)}, \,\, \text{with $C=\sqrt{2 U(\phi_i^{(m)})}$},
\ee
whose energy is $E(\phi_i^{(m)}) = L\, U(\phi_i^{(m)})$. Note that in $\mathbb{R}$, the latter solutions would have infinite energy. Finally, there is still another type of configuration in the family of solutions of (\ref{first_eq}), with nontrivial x-dependence. Those configurations are the nontrivial sphalerons $\phi_s(x)$, and unlike the constant solutions, they only arise for a specific critical length $L$ (up to multiples of $L$) due to the periodicity requirement. This sphaleron reduces to the constant solution $\phi_i^{(m)}$ in the limit $C \rightarrow \sqrt{2 U(\phi_i^{(m)})}$, where the length of the circle $L$ takes its minimum possible value $L = L_{min}$. On the other hand, it is important to emphasize that, as $C \rightarrow 0$, we have that $L \rightarrow \infty$, and the nontrivial sphaleron solution will approximate a $K\bar{K}$ pair configuration with increasing separation. This interpretation will be relevant later for the study of the sphaleron decay.

When small perturbations above the static solutions $\phi(x)$ are considered, the normal modes are obtained from the associated Schrodinger-like problem
\be\label{eq:gen_modes}
- \eta''(x) + U''(\phi(x))\eta(x) = \omega^2 \eta(x),
\ee
with the constraint $\eta(x + L) = \eta(x)$. For perturbations above constant trivial solutions $\varphi = \{\phi_i,\phi_i^{(m)}\}$, the equation gives rise to a harmonic oscillator equation, and upon imposing the periodicity condition, one obtains the allowed eigenvalues and eigenfunctions
\begin{equation}
    \eta_n(x) = 
    \begin{array}{c}
    \sin \\ \cos
    \end{array} \left(\dfrac{2\pi n x}{L}\right), \quad \omega_n^2 = U''(\varphi) + \dfrac{4\pi^2 n^2}{L^2}.
\end{equation}
The stability of the vacuum solutions $\phi_i$ implies that all the frequencies $\omega_n^2$ are positive, whereas the instability of the maximum configurations $\phi_i^{(m)}$ leads to the presence of a negative mode in the spectrum of perturbations. The equation $(\ref{eq:gen_modes})$ is in general difficult to solve analytically in the background of nontrivial sphalerons $\phi_s(x)$, and sometimes only a quasi-exactly solvable problem is obtained. A particularly interesting case occurs when $(\ref{eq:gen_modes})$ takes the following form
\begin{equation}
    - \eta''(x) - \left[ \lambda - N(N + 1) k^2 sn(x, k)^2)\right]\eta(x) = \omega^2 \eta(x),
\end{equation}
where $sn(x,k)$ denotes the Jacobi elliptic sine 
\footnotemark
\footnotetext{We define the elliptic functions throughout the text using that the incomplete integral of the first kind is expressed as $\displaystyle F(k,\psi) = \int_{0}^{\psi} \dfrac{d\,\theta}{\sqrt{1 - k\sin(\theta)}}$ \cite{Abramowitz}.}, and $\lambda, N$ and $k$ are constant parameters. This expression is identified with a Lamé differential equation, and it is well-known that if $N$ is a positive integer, the first $2N + 1$ eigenfunctions are polynomials, often referred to as Lamé polynomials \cite{Arscott-Lame}. As we will see, the $\phi^4$ and the sine-Gordon models are two examples where the spectral problem can be reduced to a Lamé equation \cite{Jiu-modes}. This equation also appears in the spectral problem of the Abelian-Higgs model in $1+1$ dimensions \cite{Brihaye-AH,Brihaye-AH2} and it is related to the spectral curve of $SU(2)$ BPS monopoles \cite{Ward-Lame,Sutcliffe-Lame}.

Finally, it is worth mentioning that, unlike the real line case, the spectrum of perturbations on the circle consists of an infinite tower of discrete square-integrable modes, that is, even the nonlocalized scattering modes have a finite norm.   

 
\section{Sphalerons on $\mathbb{S}^1$ and their internal structure}\label{sec:phi4-SG-phi6}

In this Section we review some prototypical examples of sphalerons on the circle and discuss their internal structure. The linear spectrum of perturbations about the sphalerons under study always contains a negative mode (responsible for the instability), a zero mode (due to the translational invariance of the theory) and a tower of positive modes. We will show in Sec. \ref{sec:decay} and in Sec. \ref{sec:stab} that the positive modes also play a crucial role in the sphaleron dynamics. 


\subsection{$\phi^4$ model on the circle}

Let us start with the study of the $\phi^4$ model on the circle, which is defined through the potential $U=\frac{1}{2}(1-\phi^2)^2$ and the periodicity condition $\phi(x + L) = \phi(x)$.

It well-known that the $\phi^4$ sphaleron on $\mathbb{S}^1$ can be written as follows \cite{Manton-circle}
\be\label{spha_phi4}
\phi_s(x,k) = \pm k\,a(k)\, \text{sn}( a(k)x, k^2),\quad a(k)=\sqrt{\frac{2}{1+k^2}}, \quad k\in \lbrack 0,1\rbrack.
\ee
Here, the following parametrization for the integration constant in (\ref{first_eq}) has been chosen for simplicity
\begin{equation}
    C = \dfrac{1 - k^2}{1 + k^2}.
\end{equation}
The periodicity of the Jacobi sine function imposes that the critical length $L$ at which a sphaleron may be formed is given by
\be
    L = \dfrac{4 \mathcal{K}(k^2)}{a(k)},
\ee
where $\mathcal{K}(k^2)$ is the complete elliptic integral of the first kind. When $k = 0$, the nontrivial sphaleron $(\ref{spha_phi4})$ reduces to $\phi^{(m)} = 0$, and we define $L_{min} = L|_{k = 0} = \sqrt{2} \pi$ as the smallest circle holding a sphaleron. On the other hand, the length of the circle becomes infinite when $k \rightarrow 1$, and $\phi_s(x,k)$ resembles an infinitely separated $K\bar{K}$ pair. As we have mentioned in Sec. \ref{intro}, it is also possible to have sphalerons which can be interpreted as $n$ kinks and $n$ antikinks equidistributed along the circle \cite{Manton-circle}, but in this work we will focus only on the simplest case with $n=1$. Indeed, for $n=1$, the energy of $\phi_s(x,k)$ when $L >> L_{min}$ is
\begin{equation}
    E[\phi_s] = \dfrac{8}{3} - 32\, e^{-L}, 
\end{equation}
which implies that the energy of the single sphaleron approaches exponentially the energy of the $K\bar{K}$ pair with a separation between the constituent kinks $ L/2$ \cite{Manton-energy}.

Regarding the internal structure of the sphaleron, the spectral problem can be written in the form of a Lamé-like equation
\be
    - a^2(k)\, \eta''(z) - (m - N(N + 1)\, a^2(k)\,k^2 sn(z, k^2)^2)\eta(z) = \omega^2 \eta(z),
\ee
with $z = a(k)\,x$, $m = 2$ and $N = 2$. As mentioned in Section \ref{sec:General}, for a Lamé equation with integer parameter $N$, the first $2N + 1$ eigenfunctions are the so-called Lamé polynomials. In this case, the eigenfunctions 
and their eigenfrequencies are \cite{Jiu-modes}
\be
    \begin{split}\label{eq:internal_phi4}
        \eta_{\mp} &= \text{sn}^2(z,k^2) - \dfrac{1}{3k^2}\left(1 + k^2 \pm \sqrt{1 - k^2(1 - k^2)} \right), \quad \omega_{\mp}^2 = 2\left(1 \mp 2\dfrac{\sqrt{1 - k^2(1 - k^2)}}{1 + k^2} \right) ,\\
        \eta_0 &= \text{cn}(z,k^2)\text{dn}(z,k^2), \quad \omega_0^2 = 0,\\
        \eta_1 &= \text{sn}(z,k^2)\text{cn}(z,k^2), \quad \omega_1^2 = \dfrac{6}{1 + k^2},\\
        \eta_2 &= \text{sn}(z,k^2)\text{dn}(z,k^2), \quad \omega_2^2 = \dfrac{6k^2}{1 + k^2}.
    \end{split}
\ee
where $\eta_0$ is the zero mode, $\eta_{-}$ is the negative mode and $\eta_+,\eta_1,\eta_2$ are the first positive modes. The functions $cn(x)$ and $dn(x)$ are  the Jacobi elliptic cosine and the Jacobi delta amplitude respectively. For $k$ close to $1$ (or equivalently large $L$) the modes of the sphaleron can be interpreted in terms of the individual modes of a $K\bar{K}$ pair. Of course, there is always a zero mode, responsible for the rigid translations of the sphaleron, which can be interpreted as an antisymmetric combination of the zero modes of the individual subkinks, whereas the unstable mode, $\eta_-$, can be understood as a symmetric combination. On the other hand, the positive modes $\eta_1$ and $\eta_2$ are symmetric and antisymmetric combinations of the shape mode of the individual kinks respectively. Finally, $\eta_+$ is above the mass threshold in the $K\bar{K}$ picture, therefore this is a mode genuinely related to the periodicity of $\mathbb{S}^1$. In the $k \rightarrow 1$ limit, the modes coincide with those of the $\phi^4$ kink, but they are doubly degenerate. Above $\omega_+$, there is an infinite but countable tower of extra modes, which in the large $L$ limit correspond to nonlocalized scattering states.  

Another relevant property of the $\phi^4$ sphaleron is that it has reflection symmetry. This means that the solitons on the real line satisfy $\phi(x) = - \phi(-x)$. On the circle, this property translates into $\phi(x) = - \phi(x + L/2)$. We will see in Sec. 4 that this property implies that the sphaleron has a symmetric decay along the unstable directions. 


\subsection{$\phi^6$ sphaleron on the circle} 
\label{sec:phi6}

In this section we will study the $\phi^6$ sphaleron, whose potential is $U(\phi)=\frac{1}{2}\phi^2\left(\phi^2-1\right)^2$. This is the prototypical example of a model without reflection symmetry, and in Subsec. 4.2, we will study the consequences of this lack of symmetry on the decay.

The associated static second order static equation can be integrated once to obtain
\be\label{frsit_eq_phi6}
    \phi'(x)^2-\phi^2 (\phi^2-1)^2=-C^2,
\ee
where $C$ is a constant which for the model at hand satisfies $0\leq C\leq 2/(3\sqrt{3})$. Beyond the upper bound for $C$, the periodicity condition cannot be satisfied. Upon integration, one gets
\be
    \int_{\phi(x_0)}^{\phi(x)}\frac{d\phi}{\sqrt{\phi^2(\phi^2-1)^2-C^2}} =x-x_0.
\ee

This is a hyperelliptic integral that can be computed after some algebraic manipulations \footnotemark
\footnotetext{Similar solutions were found in \cite{Sanati} in the context of kink lattices.}
\be \label{sph_phi6}
    \phi_s(x,C) = \pm\frac{1}{\sqrt{\alpha_3 - (\alpha_3-\alpha_2)\,\text{sn}(a x, b)^2}},
\ee
where
\be
    a(C) = C \sqrt{\alpha_3-\alpha_1},\quad b(C) =\sqrt{\frac{\alpha_3-\alpha_2}{\alpha_3-\alpha_1}},
\ee
and $\alpha_1, \alpha_2, \alpha_3$ are the roots of the following cubic polynomial
\be
    C^2 \alpha_i^3-\alpha_i^2+2 \alpha_i-1=0.
\ee 
\begin{figure*}[ht]
        \centering
        \includegraphics[width=0.99\textwidth]{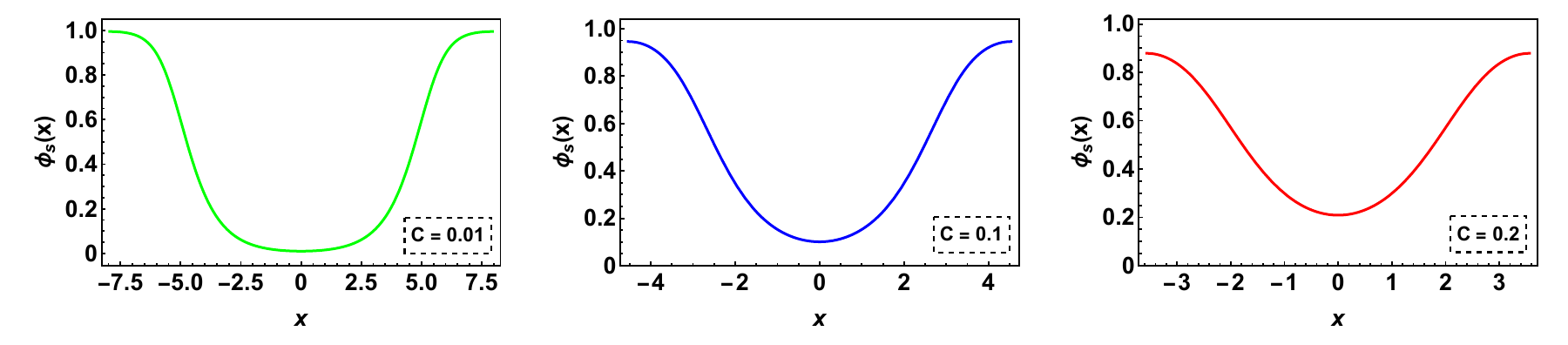}
    \caption{\small \justifying Profiles of the $\phi^6$ sphaleron $(\ref{sph_phi6})$ for different values of the parameter $C$. For $C \rightarrow 0$ the sphaleron profile resembles the $\bar{K}K$ pair in the $\phi^6$ theory on the real line.}
    \label{fig:profile_phi6}
\end{figure*}
As in the $\phi^4$ case, we will only discuss the simplest $n=1$ sphaleron, whose critical length $L$ is given by 
\be
    L = \dfrac{2 \mathcal{K}(b)}{a(C)}.
\ee
When $C = 2/(3\sqrt{3})$, the sphaleron $(\ref{sph_phi6})$ reduces to the constant unstable solution $\phi^{(m)} = 1/\sqrt{3}$, which corresponds to the smallest sphaleron, with size $L_{min} = L|_{C = 2/(3\sqrt{3})} = \sqrt{3} \pi$. The profile of the $\phi^6$ sphaleron $(\ref{sph_phi6})$ is depicted in Fig. \ref{fig:profile_phi6} for different values of the parameter $C$.

\begin{figure}
    \centering
    \includegraphics[width=0.95\linewidth]{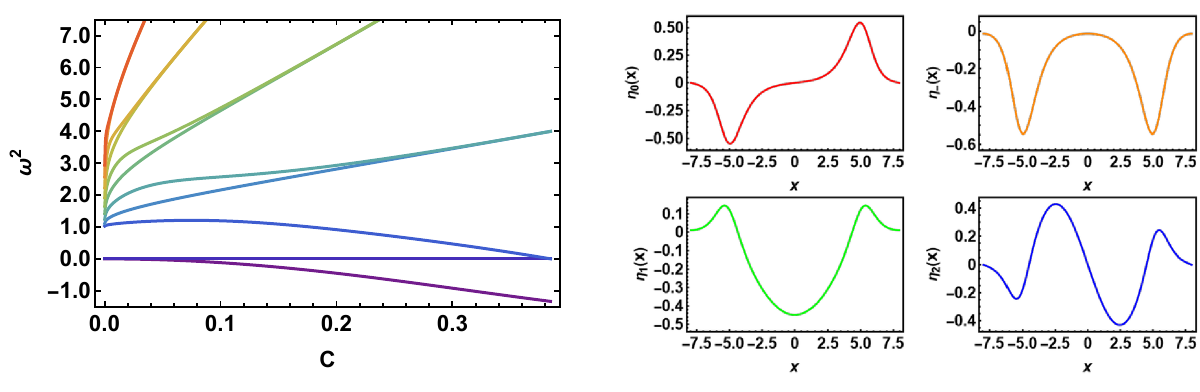}
    \caption{\small \justifying Spectrum of the first eleven linear perturbations around the sphaleron (\ref{sph_phi6}) for different values of the parameter $C$ (left). Profiles of the zero mode $\eta_0$, the unstable mode $\eta_{-}$ and the first two internal modes $\eta_{1},\eta_{2}$ hold by the $\phi^6$ sphaleron (\ref{sph_phi6}) for $C = 0.01$ (right).}
    \label{fig:phi6_eigenfunctions}
\end{figure}

The linear perturbations above the sphaleron give the following spectral problem 
\begin{equation}\label{eq:modes_phi6}
    - \eta''(x) + (1 - 12 \phi_s^2(x) + 15 \phi_s^4(x)) = \omega^2 \eta(x),
\end{equation}
where once again we must impose the constraint $\eta(x + L) = \eta(x)$. Contrary to the previous example, here we do not have a Lamé-type equation and the Schrödinger problem $(\ref{eq:modes_phi6})$ must be fully solved numerically. In Fig. \ref{fig:phi6_eigenfunctions} (left) we show the spectral flow of the modes with the model parameter $C$. We see that, except for the unstable mode, the modes degenerate in pairs in the limit of $C \rightarrow 2/(3\sqrt{3})$. Nevertheless, this degeneration breaks gradually from the lower to the higher modes as $C \rightarrow 0$. The profiles of the first eigenfunctions corresponding to the solution $(\ref{sph_phi6})$ are shown in Fig. \ref{fig:phi6_eigenfunctions} (right) for $C = 0.01$. It is easy to see that the unstable mode and the zero mode of the sphaleron can be interpreted in terms of the individual zero modes of a $K\bar{K}$ pair in the limit of $C$ close to zero: the unstable mode and the zero mode are the antisymmetric and the symmetric combinations of the zero mode of the individual subkinks respectively. Contrary to the individual $\phi^6$ kink on the real line, the $\phi^6$ sphaleron hosts internal modes. This is complete agreement with the internal mode structure of the $K\bar{K}$ pair on the real line \cite{Dorey}. 

\subsection{Sine-Gordon model on the circle} \label{sec:SG}

Let us finally consider the sine-Gordon model on the circle, where now the theory is defined through the potential $U(\phi)= 1 + \cos \phi$. As in the $\phi^4$ case, this model also has reflection symmetry. But more importantly, the sine-Gordon model is, in addition, an integrable theory on the circle. As we will show in Subsec. 4.3. this will have deep consequences in the study of the sphaleron decay.  

The nontrivial sine-Gordon sphalerons on the circle are given by the following expression \cite{Jiu-modes}
\be\label{spha_sine}
    \phi_s(x,k) = \pm 2\, \arcsin\left( k\, \text{sn}(x, k^2)\right), \quad k\in \lbrack 0,1\rbrack,
\ee
Now, the critical length $L$ reads
\begin{equation}
    L = 4 \mathcal{K}(k^2).
\end{equation}

In analogy with the $\phi^4$ case, for $k = 0$ the nontrivial sphaleron $(\ref{spha_sine})$ reduces to $\phi^{(m)} = 0$, and now $L_{min} = L|_{k = 0} = 2 \pi$ is the smallest circle holding a sphaleron. In the limit $k \rightarrow 1$, it can be shown that the energy of $(\ref{spha_sine})$ for $L >> L_{min}$ is
\begin{equation}
    E\left[\phi_s\right] = 16 - 64\, e^{-L/2}, 
\end{equation}
so that the separation between the constituent kinks is again $s = L/2$ and the energy approaches twice the sine-Gordon kink energy \cite{Rajamaran}.

Regarding the internal structure of the nontrivial sphaleron, the spectral problem can be written in the form of a Lamé-like equation
\be
    - \eta''(x) - \left(m - N(N + 1) k^2 sn(x, k^2)^2\right)\eta(x) = \omega^2 \eta(x),
\ee
where now $m = 1$ and $N = 1$. In this case, the Lamé polynomials and their frequencies are \cite{Jiu-modes}
\be
    \begin{split}
        \eta_{-} &= \text{dn}(x,k^2), \quad \omega_{-}^2 = k^2 - 1,\\
        \eta_0  &= \text{cn}(x,k^2), \quad \omega_0^2 = 0,\\
        \eta_1  &= \text{sn}(x,k^2), \quad \omega_1^2 =  k^2.
    \end{split}
\ee
The spectrum consists of a zero mode $\eta_0$, an unstable mode $\eta_-$ and a massive mode $\eta_1$ (plus an infinite tower of modes which correspond to scattering states in the large $L$ limit). Again, as $k\rightarrow 1$, the unstable mode can be interpreted as a symmetric combination of the zero modes of the component kinks. Moreover, the zero mode degenerates, whereas the massive mode becomes the threshold mode of the sine-Gordon on the real line.


\section{Sphaleron decays on the circle}\label{sec:decay}

In this section, we shall consider the decay processes of sphalerons for the theories discussed in Sec. \ref{sec:phi4-SG-phi6}. We will consider two possible initial configurations. The first one can be interpreted as the sphaleron perturbed in the direction of the unstable mode, that is,  
\begin{equation} \label{eq:lin_ansatz}
    \begin{split}
        \phi(x,0) &= \phi_s(x,k) + A\, \eta_-^{\text{\tiny $\mathcal{N}$}}(x,k),\\
        \dot{\phi}(x,0) &= A\, \omega_{-}\eta_-^{\text{\tiny $\mathcal{N}$}}(x,k),
    \end{split}
\end{equation}
where the super-index $\mathcal{N}$ reflects that the unstable mode is normalized, $A$ denotes its initial amplitude and $\omega_-$ is the corresponding associated eigenfrequency.

As discussed previously, the sphaleron resembles a kink-antikink pair in the limit $L \gg L_{min}$, so the initial condition $(\ref{eq:lin_ansatz})$ effectively represents a $K\bar{K}$ pair boosted towards each other. Therefore, the factor $A\,\omega_-$ is related to the initial velocity of the pair. However, the negative frequency $\omega_-$ is generally close to zero. This means that the initial condition $(\ref{eq:lin_ansatz})$ does not allow for ``abrupt" sphalerons decays, that is, it cannot represent high-velocity contractions of the sphaleron. For this reason, we will also explore the following initial configuration 
\begin{equation} \label{eq:rel_ansatz}
    \begin{split}
        \phi(x,0) &= \phi_{K\bar{K}}(x, x_0, v),\\
        \dot{\phi}(x,0) &= v\,\gamma(v) \phi'_{K\bar{K}}(x, x_0, v) \,,
    \end{split}
\end{equation}
where the prime denotes the derivative with respect to its argument and $\gamma$ is the Lorentz factor. Moreover, the subkink position $x_0$ is selected to ensure that the $K\bar{K}$ profile accurately fits the sphaleron profile. With this ansatz, we have direct control over the initial velocity. Note that both initial configurations (\ref{eq:lin_ansatz}) and (\ref{eq:rel_ansatz}) have different interpretations. First, the configuration (\ref{eq:lin_ansatz}) is linearly exact. This means that for $A$ small enough, the configuration converges to an exact solution. On the other hand, the configuration (\ref{eq:rel_ansatz}) becomes exact for $L \gg L_{min}$. In this situation the sphaleron is big enough and the component kinks can be boosted independently as if they were a $K\bar{K}$ pair on the real line. As a consequence, this initial configuration does not make sense if the distance between the component kinks is on the order of the kink size itself.   

Throughout the following sections, time evolutions have been performed using a second-order finite difference scheme in both space and time with periodic boundary conditions. The simulation box is the interval $-L/2 \leq x \leq L/2$ with the identification $-L/2\sim L/2$, with a typical number of grid points $n_{space} = 1000$, grid spacing $\Delta x = L/n_{space}$ and time step $\Delta t = \Delta x/4$. 


\subsection{Sphaleron decay in the $\phi^4$ model}

\begin{figure*}[ht]
    \centering
    \begin{subfigure}[t]{0.5\textwidth}
        \centering
        \includegraphics[width=0.675\textwidth]{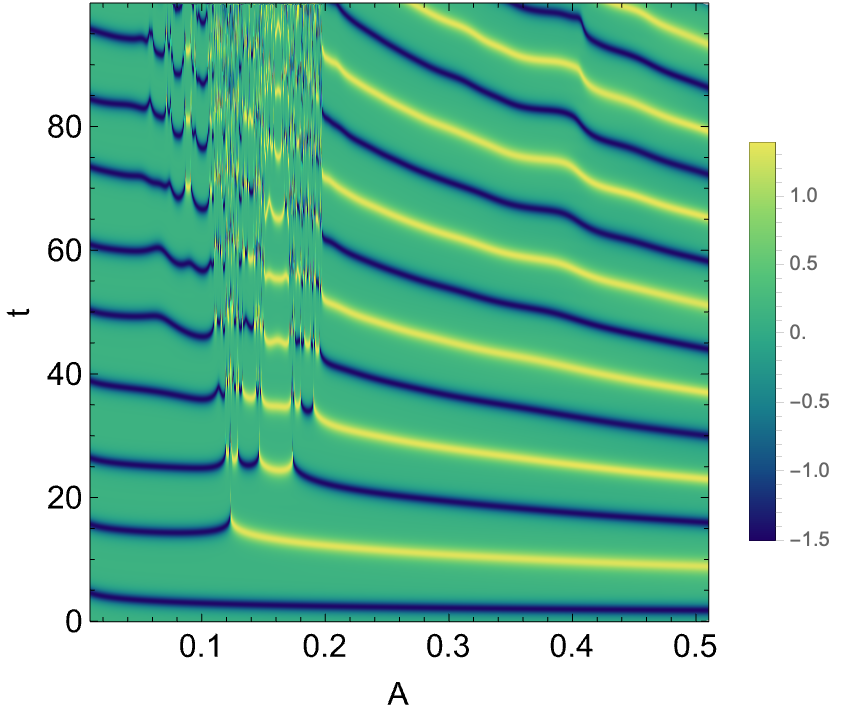}
        \caption{\small $\phi_s(L/4,t)$ for $k = 0.1$.}
        \label{fig:decay_phi4_k_small}
    \end{subfigure}%
    \hfill
    \begin{subfigure}[t]{0.5\textwidth}
        \centering
        \includegraphics[width=0.7\textwidth]{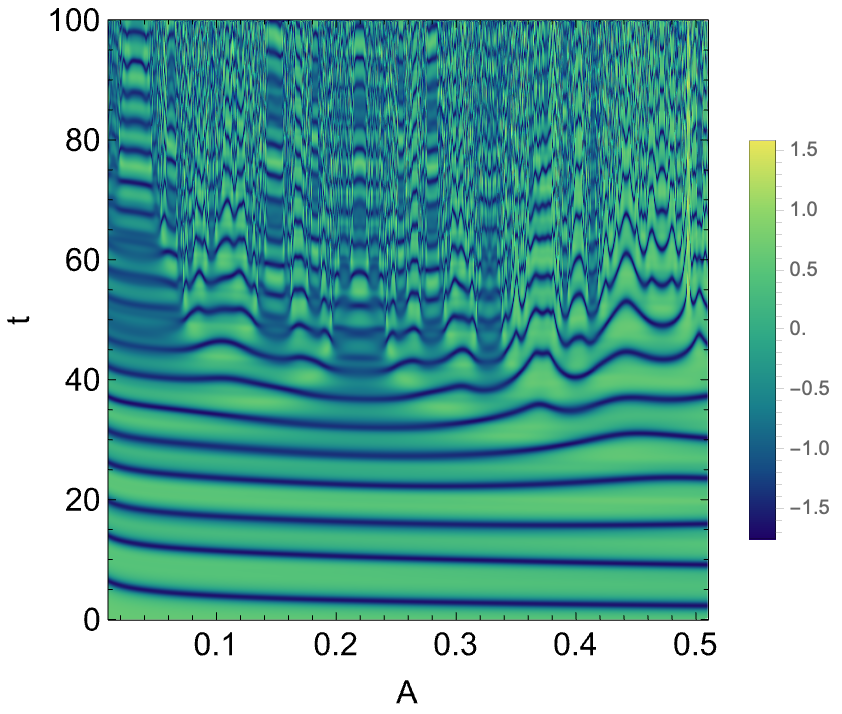}
        \caption{\small $\phi_s(L/4,t)$ for $k = 0.5$.}
        \label{fig:decay_phi4_k_moderate}
    \end{subfigure}%
    \hfill 
    \begin{subfigure}[t]{0.5\textwidth}
        \centering
        \includegraphics[width=0.7\textwidth]{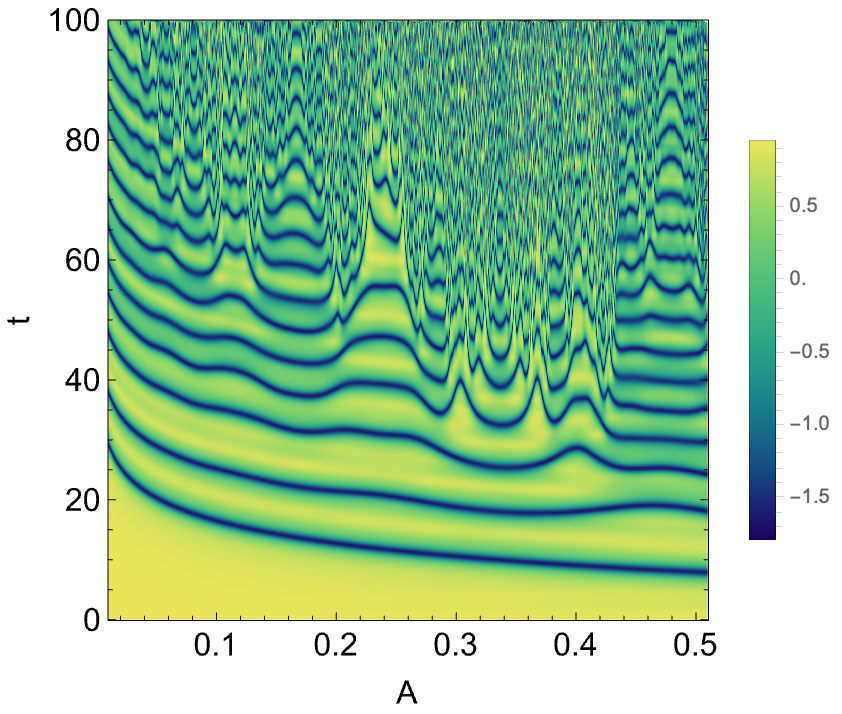}
        \caption{\small $\phi_s(L/4,t)$ for $k = 0.9$.}
        \label{fig:decay_phi4_k_big}
    \end{subfigure}

    \caption{\small \justifying Decay of the $\phi^4$ sphaleron for different initial amplitudes of the unstable mode $A$ using the initial condition $(\ref{eq:lin_ansatz})$.}
    \label{fig:decay_phi4}
\end{figure*}

We begin our analysis by studying the sphaleron decay in $\phi^4$ for small values of $k$ with the initial condition $(\ref{eq:lin_ansatz})$. We have considered three representative values for $k$: $k = 0.1,\, 0.5$ and $0.9$. The corresponding evolution of the field for different initial values of the unstable amplitude is illustrated in Fig. \ref{fig:decay_phi4}. The color palette accounts for the field value at $x = L/4$, which corresponds to the position on the circle where the subkinks scatter.

For $k = 0.1$ (see Fig. \ref{fig:decay_phi4_k_small}), three regions can be distinguished: In the region $A \lessapprox 0.06$, the sphaleron initially decays along the negative direction due to our choice of sign and reaches a minimum. At that point, the instant configuration bounces back and approaches the initial profile. Then, the decay along the same direction repeats almost periodically. Now, in the region $0.06 \lessapprox A \lessapprox 0.2$, after the sphaleron decay, the system is again close to the unexcited sphaleron. Then, the system chaotically decays either along the positive or negative direction. Finally, for $A \gtrapprox 0.2$, the sphaleron exhibits an oscillatory behavior alternating between both vacua. For $k = 0.5$ (see Fig. \ref{fig:decay_phi4_k_moderate}) the dynamics is more chaotic, and for all initial amplitudes of the unstable mode the decay only occurs along one fixed direction, and after each bounce the configuration at the turning point gets further from the initial field configuration. As $k \rightarrow 1$, the sphaleron decay resembles the $K\bar{K}$ scattering (see Fig. \ref{fig:decay_phi4_k_big} for $k = 0.9$). The component subkinks approach each other and collide forming an oscillon state. But, for some intermediate values of $A\,\, (0.2<A<0.3) $, the oscillon period grows, resembling the back-scattering of a $K\bar{K}$ pair that scatters on the real line. 

For the initial configuration (\ref{eq:lin_ansatz}) and $k$ closer to $1$, we expect a behavior similar to Fig. \ref{fig:decay_phi4_k_big}. This is because, as $k$ grows to $1$, $\omega_{-}$ approaches 0. This means that the kinetic contribution to the initial configuration vanishes and (\ref{eq:lin_ansatz}) simply represents a $K\bar{K}$ pair at a large distance attracted by the static intersolitonic force. This effectively means that the initial configuration corresponds to a $K\bar{K}$ pair with a vanishing center of mass velocity. As a consequence, the expected behavior should correspond to a $K\bar{K}$ scattering at very low initial velocities.  

\begin{figure*}[ht]
    \centering
    \begin{subfigure}[t]{0.5\textwidth}
        \centering
        \includegraphics[width=0.89\textwidth]{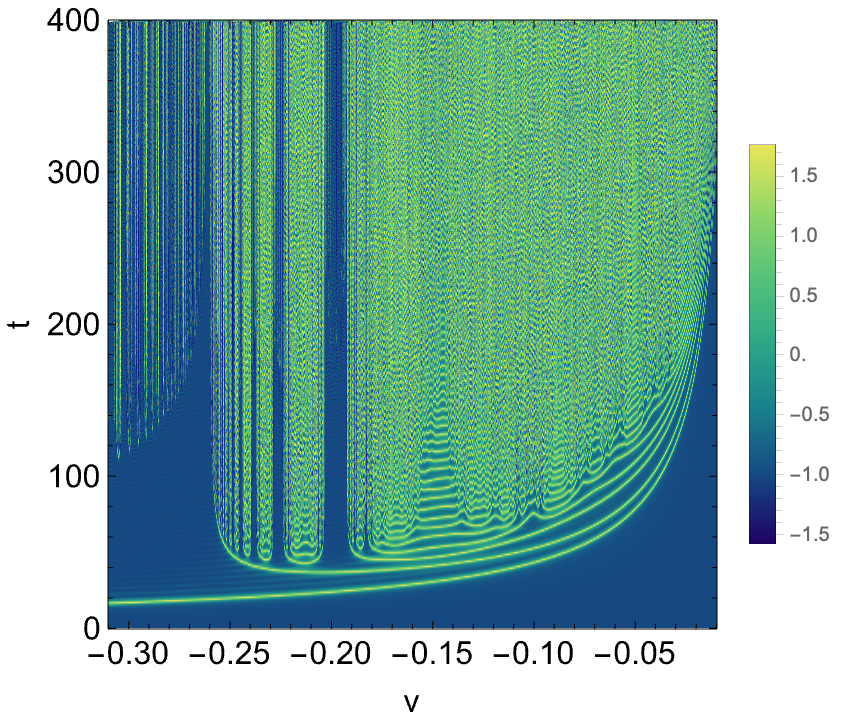}
        \caption{\small $\phi_s(-L/4,t)$ in the $\bar{K}K$ scattering.}
        \label{fig:decay_phi4_positive_rel}
    \end{subfigure}%
    \begin{subfigure}[t]{0.5\textwidth}
        \centering
        \includegraphics[width=0.9\textwidth]{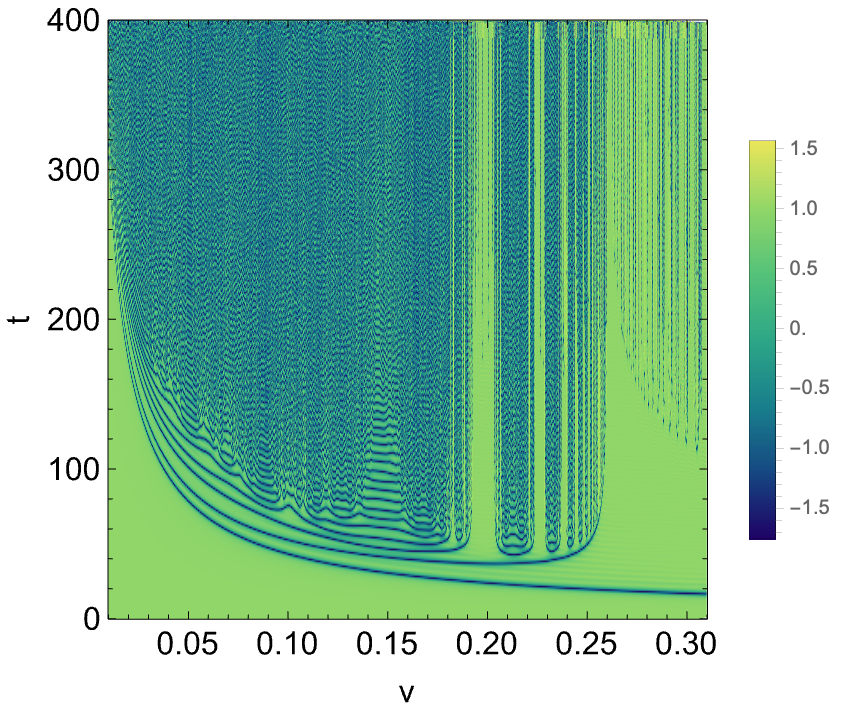}
        \caption{\small $\phi_s(L/4,t)$ in the $K\bar{K}$ scattering.}
        \label{fig:decay_phi4_negative_rel}
    \end{subfigure}

    \caption{\small \justifying $\phi^4$ sphaleron decay with the initial condition (\ref{eq:rel_phi4}) for $k = 0.9999$. The bounce windows and the critical velocity of the $\phi^4$ $K\bar{K}$ (and $\bar{K}K$) pair scattering are recovered.}
    \label{fig:phi4_rel}
\end{figure*}

In order to study a rapid collapse of the sphaleron, we can go to a regime where $k$ is close to 1 and use the initial configuration $(\ref{eq:rel_ansatz})$. For the $\phi^4$ model, this initial configuration would read as
\begin{equation} \label{eq:rel_phi4}
    \begin{split}
        \phi(x,0) &= \tanh(\gamma \,x) - \tanh(\gamma \,(x - L/2)) - \tanh(\gamma \,(x + L/2)),\\
        \dot{\phi}(x,0) &= v\, \gamma\left(\sech^2(\gamma \,x) + \sech^2(\gamma \,(x - L/2)) + \sech^2(\gamma \,(x + L/2))\right).
    \end{split}
\end{equation}
As explained before, this corresponds to a $K\bar{K}$ pair located at diametrically opposite positions on the circle and boosted towards each other. We have chosen the value $k = 0.9999$ (see Fig. \ref{fig:phi4_rel}). We observe that the behavior of the collapsing sphaleron corresponds exactly to the $K\bar{K}$ fractal scattering pattern on the real line (see, for example, \cite{Campbell}). Of course, due to the periodicity of the base space, nothing flies away, and at times much larger than the base space size, the outgoing subkinks scatter again. This can be clearly seen in the region above the critical velocity $v>0.25$. Note also that for negative velocities, i.e., when the sphaleron collapses to the other vacuum, the fractal pattern remains exactly the same. This is, as we have mentioned, due to the reflection symmetry on the circle $\phi_s(x)=-\phi_s(x+L/2)$. 

\subsection{Decay of the $\phi^6$ sphaleron on the circle}\label{sec:decay_phi6}

In this section, we shall discuss the sphaleron decay in $\phi^6$ theory. Due to the lack of reflection symmetry, the decay must be analyzed separately along each unstable direction. For this reason, we will use the initial conditions $(\ref{eq:lin_ansatz})$ and $(\ref{eq:rel_ansatz})$ for both positive and negative amplitudes of the unstable mode and velocities.

\begin{figure*}[ht]
    \centering
    \begin{subfigure}[t]{0.45\textwidth}
        \centering
        \includegraphics[width=0.97\textwidth]{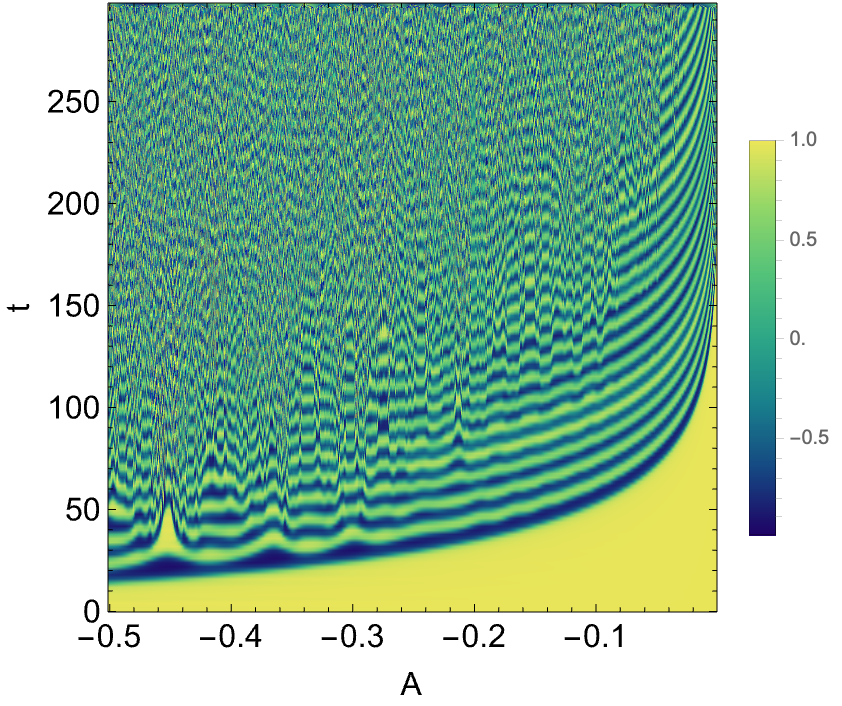}
        \caption{\small $\phi_s(L/2,t)$ for $C= 0.01$.}
        \label{ffig:decay_phi6_negative_moderateC}
    \end{subfigure}
   \begin{subfigure}[t]{0.45\textwidth}
        \centering
        \includegraphics[width=0.99\textwidth]{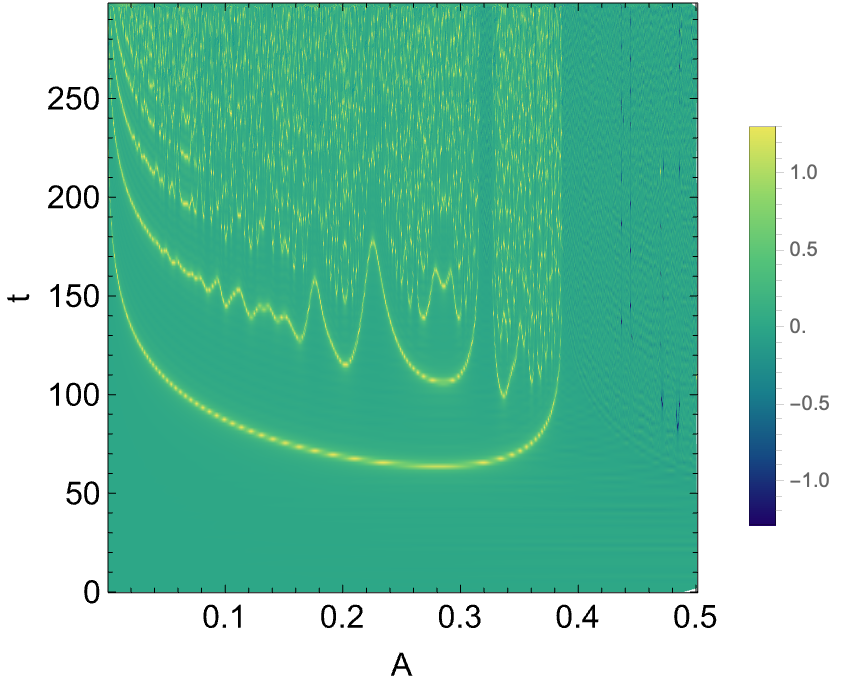}
        \caption{\small $\phi_s(0,t)$ for $C = 0.01$.}
        \label{fig:decay_phi6_positive_smallC}
    \end{subfigure}%
   \caption{\small \justifying Decay of the $\phi^6$ sphaleron for negative values of the unstable amplitude (left panel) and for positive values of the unstable amplitude $A$ (right panel) using the initial condition $(\ref{eq:lin_ansatz})$.}
    \label{fig:decay_phi6_linear_neg}
\end{figure*}

We will start the study using the initial configuration given by (\ref{eq:lin_ansatz}) with the choice of the integration constant $C=0.01$. First, for negative values of the amplitude, the sphaleron collapses and produces an oscillon (see Fig. \ref{ffig:decay_phi6_negative_moderateC}). A much richer structure arises for positive values of the amplitude, Fig. \ref{fig:decay_phi6_positive_smallC}. In this case, the sphaleron decay is more chaotic. It forms a bion state for some range of the amplitudes, where the sphaleron bounces several times before it forms an oscillon. In a small window, $A\sim 0.32$, the sphaleron only bounces once, and the component kinks move apart from $x=0$ to the opposite side of the circle, where an oscillon state is formed. More intriguing is what happens at $A\sim 0.39$. At this point, it seems that the sphaleron freezes at some fixed position for a long time. This behavior has been observed for other values of the parameter $C$, showing a variable number of small windows before the sphaleron stops. 

The different behaviors observed at positive and negative amplitudes can be traced back to the internal structure of the sphaleron. In the negative amplitude case, the sphaleron resembles the $(0, 1)+(1, 0)$ $K\bar{K}$ collision on the real line, i.e., a collision between a kink interpolating between the vacua $0 \rightarrow 1$ and an antikink interpolating between the vacua $1 \rightarrow 0$. It is well-known that this configuration does not have internal modes \cite{Dorey}. 

The sphaleron configuration hosts bound modes, but their support is localized in the region outside the collapse (see Fig. \ref{fig:phi6_eigenfunctions}). This effectively implies that the energy cannot be stored in the bound modes during the collapse and the fractal structure does not appear. For positive amplitudes, the situation is opposite. Now, the internal modes have support in the region of collapse. During decay, they can be excited and the resonant energy transfer mechanism takes place.

In order to support this hypothesis, we perform a slightly different experiment. We start now with the initial configuration given by $(\ref{eq:rel_ansatz})$ for $C=0.0001$. To be precise, we use the following $K\bar{K}$ ansatz 
\begin{equation} \label{eq:phi6_rel}
    \begin{split}
        \phi(x,0) &= \sqrt{\dfrac{1 - \tanh(\gamma(x + x0))}{2}} + \sqrt{\dfrac{1 + \tanh(\gamma(x - x0))}{2}},\\
        \dot{\phi}(x,0) &= v\, \gamma\left(\dfrac{\sech^2(\gamma(x + x0))}{2\sqrt{2}(1 - \tanh(\gamma(x + x0)))} + \dfrac{\sech^2(\gamma(x - x0))}{2\sqrt{2}(1 + \tanh(\gamma(x - x0)))}\right),
    \end{split}
\end{equation}
with $x_0 = 9.89$.

\begin{figure*}[ht]
    \centering
    \begin{subfigure}[t]{0.5\textwidth}
        \centering
        \includegraphics[width=0.91\textwidth]{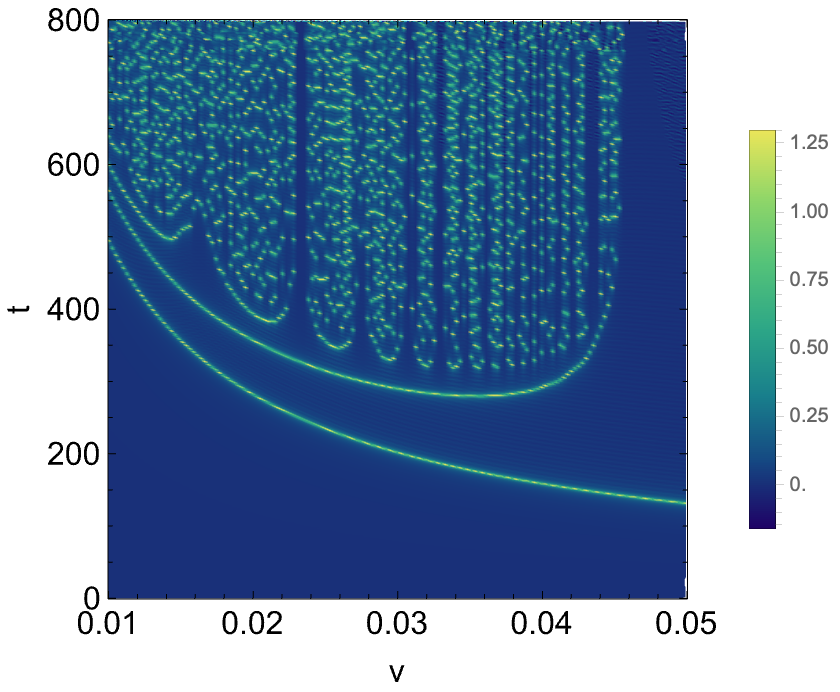}
        \caption{\small $\phi_s(0,t)$ in the $\bar{K}K$ scattering.}
        \label{fig:decay_phi6_positive_rel}
    \end{subfigure}%
    \begin{subfigure}[t]{0.5\textwidth}
        \centering
        \includegraphics[width=0.88\textwidth]{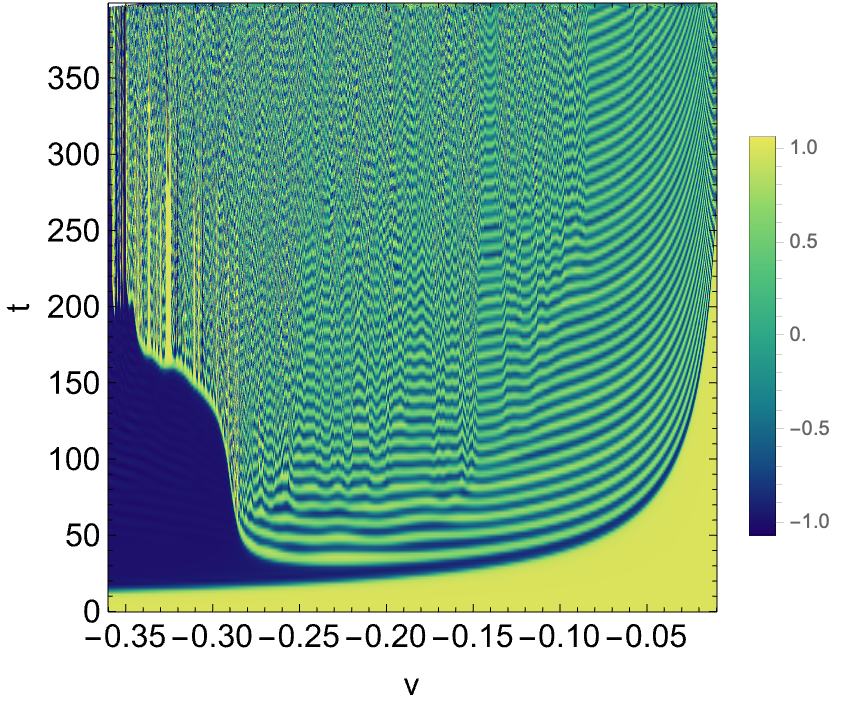}
        \caption{\small $\phi_s(L/2,t)$ in the $K\bar{K}$ scattering.}
        \label{fig:decay_phi6_negative_rel}
    \end{subfigure}

    \caption{\small \justifying Sphaleron decay in the $\phi^6$ model for positive (left panel) and negative (right panel) velocities when $C = 0.0001$ and for the initial condition $(\ref{eq:phi6_rel})$. The bounce windows and the critical velocity of the $\phi^6$ antikink-kink pair scattering is recovered. The critical velocity in the kink-antikink pair scattering is also captured.}
    \label{fig:phi6_rel}
\end{figure*}

We have performed simulations for the sphaleron decay, both for negative and positive initial velocities, corresponding to $ (0, 1)+(1, 0)$ $K\bar{K}$ and $ (1, 0)+(0, 1)$ $\bar{K}K$ configurations, respectively. The time evolution for different velocities is shown in Fig. \ref{fig:phi6_rel}. The left panel shows $ (1, 0)+(0, 1)$ $\bar{K}K$ collision, where the characteristic fractal pattern is clearly observed \cite{Dorey}. The critical velocity and bounce windows are accurately determined. The right panel shows the $ (0, 1)+(1, 0)$ $K\bar{K}$ collision, where the fractal pattern is absent; instead, only the decay into a bion is seen, along with the critical velocity at which the pair scatters back inelastically. 

The behavior observed in Fig. \ref{fig:decay_phi6_positive_smallC} at $A\sim 0.39$ deserves a separate explanation. As we discussed above, it seems that at this point the sphaleron freezes at some point and the collapse stops. This phenomenon is also related to the excitation of the internal modes but in a different way. When the unstable mode is excited, the positive bound modes get excited at quadratic order. As the amplitude of the negative mode grows, the sphaleron contracts and the bound modes increase their frequency. However, this increase in frequency generates an effective force that opposes the contraction of the sphaleron, and an equilibrium point takes place. This phenomenon was observed by the authors in the context of false vacuum sphalerons \cite{Sergio-phi6}. We have left for Sec. \ref{sec:stab} a detailed explanation of this mechanism. 


\subsection{Sphaleron decay in the Sine-Gordon model}

In Subsec. 3.3, we introduced the sine-Gordon sphalerons (\ref{spha_sine}). In this section, we will study the decay of these solutions after a small perturbation in the unstable direction. 

\begin{figure*}[ht]
    \centering    
    \begin{subfigure}[t]{0.4\textwidth}
        \centering
        \includegraphics[width=0.9\textwidth]{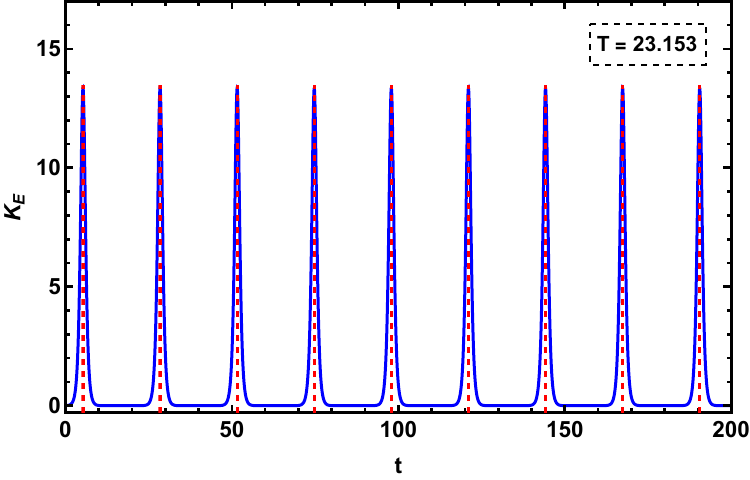}
        \caption{\small Kinetic energy for $A = 0.1$.}
        \label{fig:KE_k0.5_A0.1}
    \end{subfigure}%
    \hspace{0.5cm}
    \begin{subfigure}[t]{0.4\textwidth}
        \centering
        \includegraphics[width=0.9\textwidth]{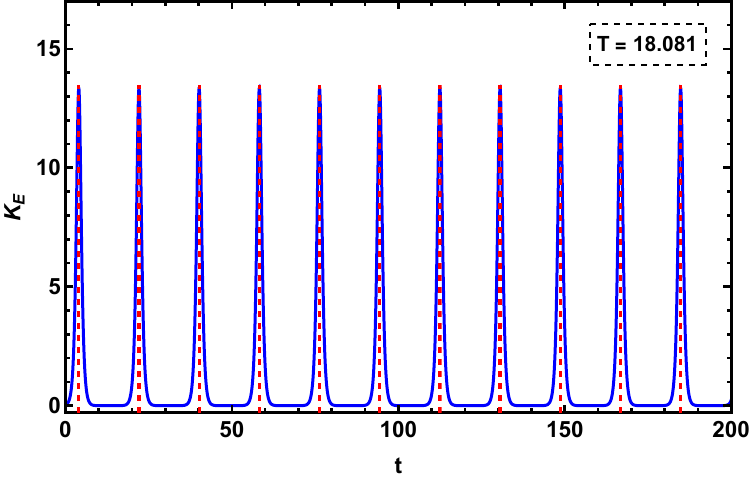}
        \caption{\small Kinetic energy for $A = 0.3$.}
        \label{fig:KE_k0.5_A0.3}
    \end{subfigure} \vspace{1em}
    \begin{subfigure}[t]{0.4\textwidth}
        \centering
        \includegraphics[width=0.924\textwidth]{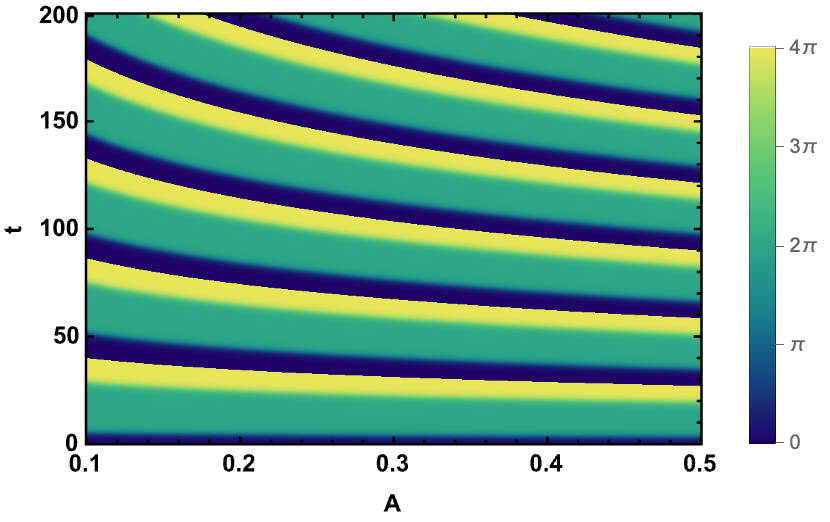}
        \caption{\small \justifying $\phi(0,t)$ for different initial values of the unstable amplitude $A$.}
        \label{fig:DecaySG}
    \end{subfigure}%
    \hspace{0.25cm}
    \begin{subfigure}[t]{0.4\textwidth}
        \centering
        \includegraphics[width=0.9\textwidth]{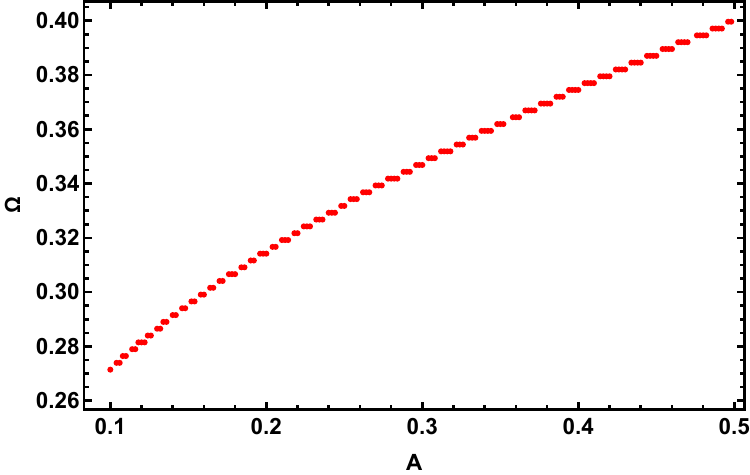}
        \caption{\small \justifying Kinetic energy periodicity for different initial unstable amplitudes $A$.}
        \label{fig:KE_k0.5_All}
    \end{subfigure}
 
    \caption{\small \justifying Upper panel: Kinetic energy as a function of time for two selected cases, illustrating the characteristic oscillatory behavior of the sine-Gordon sphaleron. Lower left panel: Evolution of the sphaleron profile at $x = 0$ (mod $4\pi$). Lower right panel: Temporal angular frequency of the kinetic energy during the decay of the sine-Gordon sphaleron for different values of the initial amplitude of the unstable mode $A$. All plots assume $k = 0.5$.}
    \label{fig:KE_k0.5}
\end{figure*}

In our next numerical experiment, we choose again the initial configurations (\ref{eq:lin_ansatz}) for a sphaleron of the form (\ref{spha_sine}). In Figs. \ref{fig:KE_k0.5_A0.1} and \ref{fig:KE_k0.5_A0.3} we plot the kinetic energy as a function of time for different values of the parameter $k$ (see also Fig. \ref{fig:DecaySG} for the evolution of the sphaleron profile at $x=0$). We observe a series of periodic peaks increasing in frequency as the amplitude of the unstable mode increases. As we have explained above, for an initial configuration of the form (\ref{eq:lin_ansatz}), the initial kinetic energy is very small. As the configuration evolves in the unstable direction, the kinetic energy increases rapidly. At this point, the sphaleron transitions to an (almost) antisphaleron configuration centered at  the nearest maximum ($\phi_{max}=\pm \pi$) of potential, with approximately zero kinetic energy, and stays there for a while until it goes back to the initial sphaleron configuration. The direction of decay is decided again depending on the sign of the unstable mode amplitude.  This oscillation repeats periodically with a fixed angular frequency $\Omega$.

Of course, this regular behavior is intimately related to the integrability of the model. Interestingly, the sphaleron decay in the sine-Gordon model can be related to an analytical time-periodic solution of the form
\be\label{spha_sine_time}
    \phi_s(x,t) = \pm \left( 4\, \arctan\left( A\, \text{dn}(z, m_1)\,\text{sn}(f\, t, m_2)\right) \pm \pi\right),
\ee
where
\bea
z&=& \beta(x - L/4), \quad L = \frac{2}{\beta}K(m_1), \quad \beta = f A.\\
m_1&=&1+\frac{1}{A^2}-\frac{1}{\beta^2(1+A^2)}, \quad m_2 = \frac{A^2}{f^2 (1+A^2)}-A^2.
\eea
A solution of the form (\ref{spha_sine_time}) is usually called breather oscillation in the literature \cite{costabile, marchesoni}. The correspondence between solutions of the form (\ref{spha_sine_time}) and sphaleron decays can be done simply by imposing the appropriate constraints on the $A$ and $f$ parameters. For a sphaleron of length $L$ and observed frequency $\Omega$ (obtained numerically) the analytical solution (\ref{spha_sine_time}) must satisfy the same time and space periodicities. This translates into the following relations 
\bea
L&=&\frac{2}{f A}\mathcal{K}(m_1), \\
\Omega&=& \frac{\pi f}{2\mathcal{K}(m_2)}.
\eea
The observed periodicity is a crucial (and expected) distinction of this model compared to previous ones, enabling analytical knowledge of the sphaleron decay at all times. This may provide an interesting framework for testing sphaleron properties. 

 
 \section{Sphaleron stabilization and internal modes}\label{sec:stab}

In Sec. 4.2 we argued that the excitation of the internal modes may act as an obstacle during the sphaleron decay, leading to a quasi-static solution. This phenomenon can be seen as a sort of stabilization for the sphaleron triggered by the internal positive modes. In this section, we give a concise explanation behind the mechanism.

\begin{figure}
    \centering
    \includegraphics[width=0.75\linewidth]{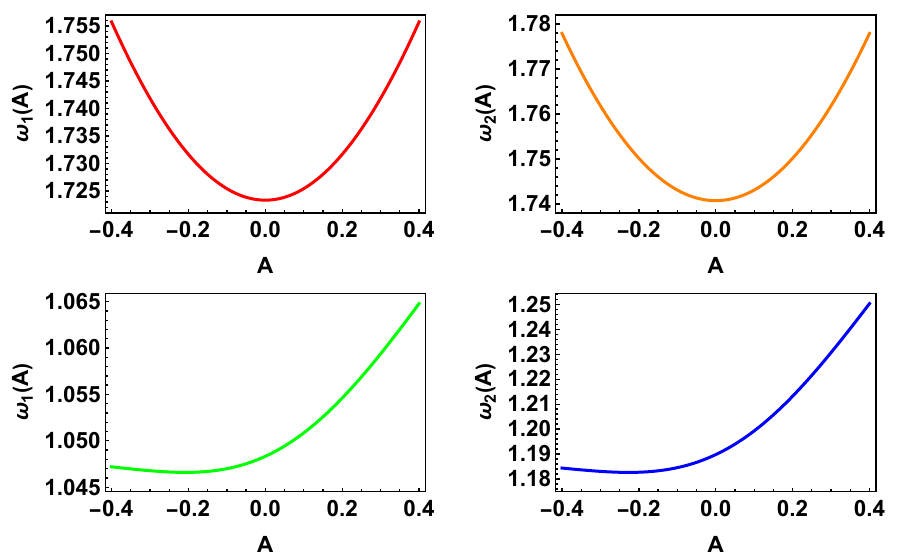}
    \caption{\small \justifying Spectral flow of the first two positive internal modes around the $\phi^4$ sphaleron for $k = 0.99$ (Upper panel) and around the $\phi^6$ sphaleron for $C = 0.001$ (Lower panel). The amplitude $A$ denotes the strength of the corresponding unstable mode.}
    \label{fig:spectralFlow_phi4_phi6}
\end{figure}

The qualitative idea is the following. The initial excitation of the unstable mode leads to an effective change in the size of the sphaleron. As the sphaleron leaves its unstable equilibrium state, it feels a static attractive force. On the other hand, the linear modes of the sphaleron have a fixed position in the linear spectrum. As the sphaleron starts to collapse, the modes do not disappear but move through the spectrum, changing their shape and frequency. Generically, this spectral flow generates an effective force given by
\begin{equation}\label{force_rep}
    F_{s} = - \dfrac{1}{2}\dfrac{d\omega^2(A)}{dA}B^2,
\end{equation}
where $\omega(A)$ is the frequency of the positive mode depending on the sphaleron deformation and B is its amplitude. This force is repulsive since $\omega$ grows as the sphaleron collapses (see Fig. \ref{fig:spectralFlow_phi4_phi6}). 
The balance between these two forces leads to the critical value of the amplitude of the shape mode $B_c$ for which the sphaleron does not decay. This idea of balance between forces can be accomplished by means of a perturbative calculation \cite{adam-quasi, Sergio-phi6}. Let us assume that our solution can be expanded in terms of the modes in the following way
\be\label{ansatz:stab}
\phi(x,t)=\phi_s(x) - A_c\, \eta_{-}(x) + A_{-}(t) \,\eta_{-}(x) + \sum_i A_i\,\phi_i^{(1)}(x,t) + \sum_i A^2_i\phi_i^{(2)}(x,t),
\ee
where $A_c$ is a constant at which the sphaleron stabilizes, $A_{-}$ is a possible amplitude for the negative mode and $\phi^{(i)}$ are the linear and quadratic corrections to the positive modes. By substituting (\ref{ansatz:stab}) into the corresponding field equations and projecting onto the unstable mode, we get the following expression
\be \label{eq:expansion_Aformula}
    \ddot{A}_{-}(t) + \omega_{-}^2\,A_{-}(t) = A_c\,\omega_{-}^2 - \frac{1}{2}\int_{\mathbb{S}^1} U'''(\phi_s)\,\eta_{-}(x)\left(\sum_i A_i \,\phi_i^{(1)}(x,t)\right)^2\,dx,
\ee
at first order in $A_-$ and second order in $A_i$. The initial excitation of the unstable mode serves as a quadratic source for the linear corrections to the positive modes $\phi_i^{(1)} = \eta_i(x)\cos(\omega_i t)$ due to the nonlinearity of the model. The amplitude at which these positive modes are excited by the critical amplitude of the unstable mode $A_c$ is given, at the lowest order, by the following expression
\begin{equation}
A_i = A_{c}^2\,\dfrac{\gamma_i}{\omega_i^2},
\end{equation}
 with 
\begin{eqnarray}\label{gamma:eq}
\gamma_i &=& \dfrac{1}{2} \int_{\mathbb{S}^1}U^{''}(\phi_s)\,\eta_{-}^2(x)\,\eta_{i}(x)\, dx\,.
\end{eqnarray} 
Therefore, the right-hand side of $(\ref{eq:expansion_Aformula})$ is given in terms of constant and oscillatory terms of the form $\cos(\omega_i t)\cos(\omega_j t)$. To avoid the exponential growth of $A_{-}$ and stabilize the sphaleron, we impose that the nonoscillatory contribution vanishes. This condition gives precisely the critical value of the amplitude
\begin{equation}\label{eq:crit}
    A_c = \left(\frac{4\,\omega_{-}^2}{\sum\limits_i \frac{\gamma_i^2 \delta_i}{\omega_i^4}}\right)^{1/3}\,,
\end{equation}
where
\begin{eqnarray}
    \delta_i = \int_{\mathbb{S}^1} U'''(\phi_s)\,\eta_{-}(x)\,\eta_i(x)^2\,dx\,.
\end{eqnarray}

\begin{figure*}[ht]
    \centering
    \begin{subfigure}[t]{0.5\textwidth}
        \centering
        \includegraphics[width=0.99\textwidth]{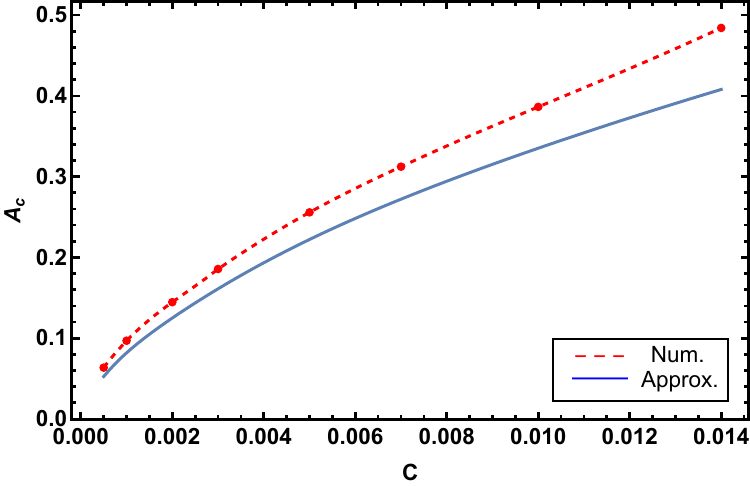}
        \caption{\small \justifying Numerical and analytical critical amplitude for the $\phi^6$ model.}
        \label{fig:ComparionAc}
    \end{subfigure} \vspace{1em}  
    \begin{subfigure}[t]{0.5\textwidth}
        \centering
        \includegraphics[width=0.895\textwidth]{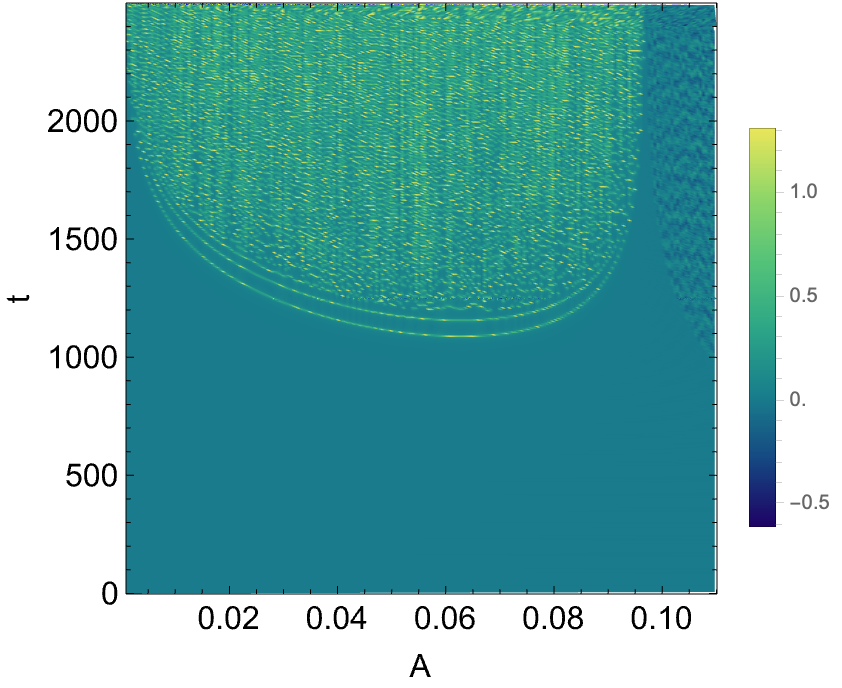}
        \caption{\small $\phi_s(0,t)$ for $C = 0.001$. }
        \label{fig:Stabilization_phi6_A0.001}
    \end{subfigure}%
    \begin{subfigure}[t]{0.5\textwidth}
        \centering
        \includegraphics[width=0.88\textwidth]{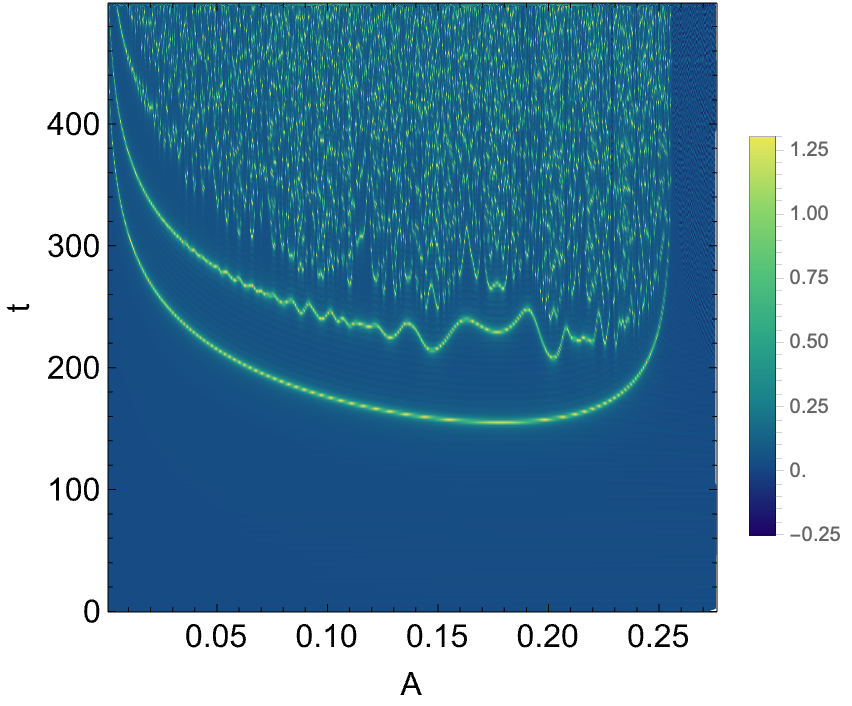}
        \caption{\small $\phi_s(0,t)$ for $C = 0.005$.}
        \label{fig:Stabilization_phi6_A0.005}
    \end{subfigure}

    \caption{\small \justifying Upper panel: Comparison between the numerical critical amplitude and the analytical prediction given by equation $(\ref{eq:crit})$ for the $\phi^6$ model for different values of the model parameter $C$. Lower panel: Stabilization of the $\phi^6$ sphaleron $(\ref{sph_phi6})$ with $C = 0.001$  and $C = 0.005$. The amplitude $A$ accounts for the initial excitation of the unstable mode.}
    \label{fig:phi6_stabilization}
\end{figure*}

A comparison between (\ref{eq:crit}) and full-numerical results is shown in Fig. \ref{fig:ComparionAc} for the $\phi^6$ model. Two representatives examples of the stabilization in $\phi^6$ can be found in Figs. \ref{fig:Stabilization_phi6_A0.001} and \ref{fig:Stabilization_phi6_A0.005}. The agreement between the analytical formula and full numerics is good, although at large critical amplitudes, higher-order perturbative effects have to be taken into account. This should be responsible for the small deviations observed in Fig. \ref{fig:ComparionAc} as $A_c$ grows. 

\begin{figure*}[ht]
    \centering
    \begin{subfigure}[t]{0.5\textwidth}
        \centering
        \includegraphics[width=0.855\textwidth]{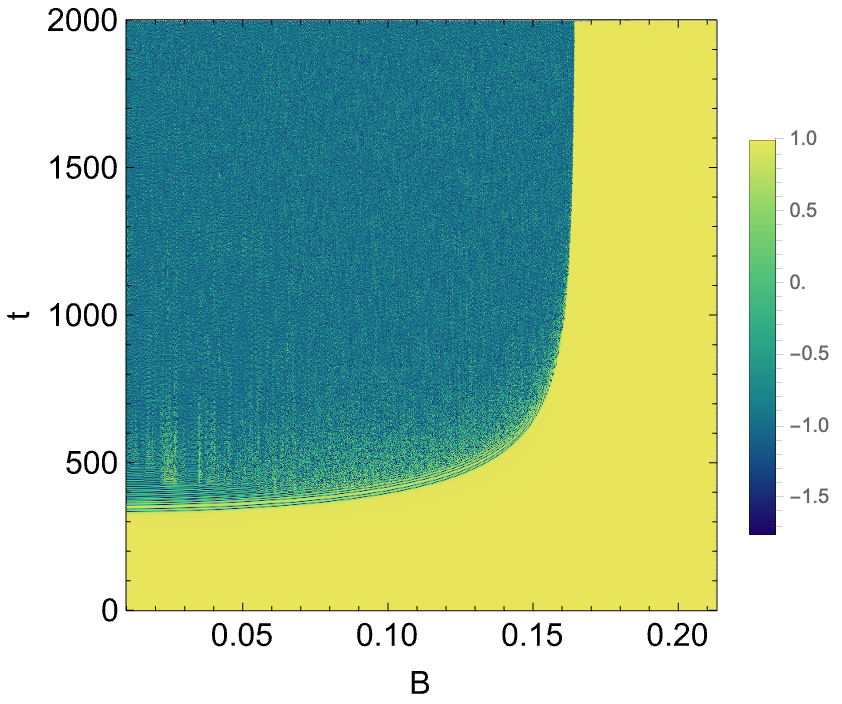}
        \caption{\small $\phi_s(L/4,t)$ for $A = 0.05$.}
        \label{fig:Stabilization_phi4_A0.05}
    \end{subfigure}%
    \hspace{-0.15cm}
    \begin{subfigure}[t]{0.5\textwidth}
        \centering
        \includegraphics[width=0.865\textwidth]{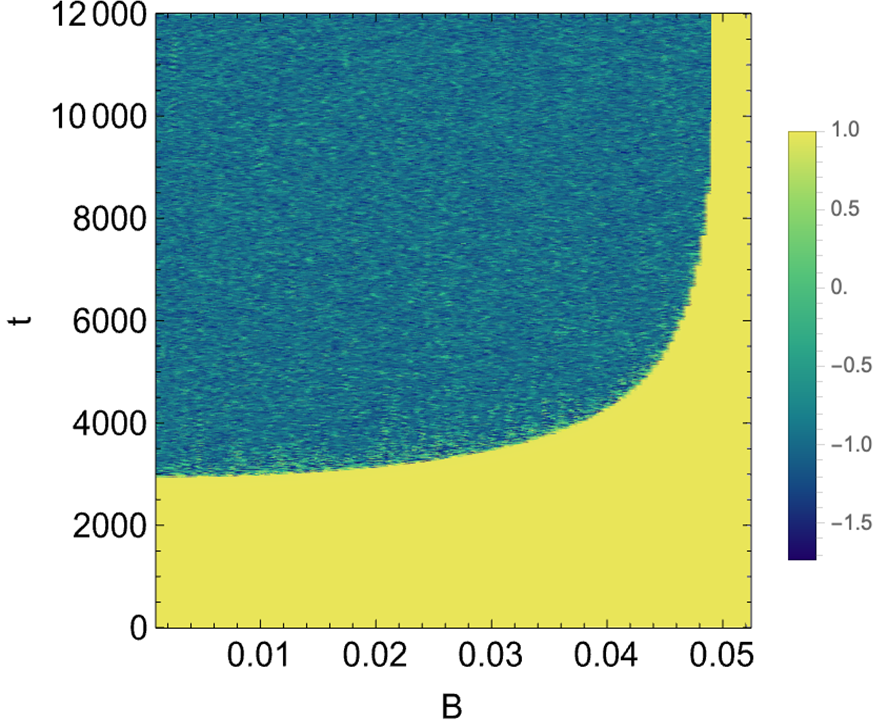}
        \caption{\small $\phi_s(L/4,t)$ for $A = 0.01$. }
        \label{fig:Stabilization_phi4_A0.01}
    \end{subfigure}%

    \caption{\small \justifying Dynamical stabilization of the $\phi^4$ sphaleron with $k = 0.99$  (left panel) and $k = 0.999$ (right panel). The initial unstable amplitude is given by $A$ and the amplitude $B$ accounts for the initial excitation of the first positive internal mode $\eta_1$ around the sphaleron $(\ref{eq:internal_phi4})$.}
    \label{fig:phi4_stabilization}
\end{figure*}

In $\phi^4$ theory we observe a similar behavior (see Fig.  \ref{fig:phi4_stabilization}). In this case, the sphaleron cannot be stabilized simply by exciting the unstable mode. A quick look at (\ref{gamma:eq}) shows that, due to the reflection symmetry, the positive modes cannot be excited at quadratic order by the excitation of the unstable mode. This implies that, in order to accomplish the stabilization, the positive modes have to be excited independently. In addition, it seems that the balance between forces is a higher-order effect. For instance, the integral in the r.h.s. of (\ref{eq:expansion_Aformula}) vanishes identically in the $\phi^4$ model, implying that the balance cannot be a quadratic effect. A precise prediction of the critical amplitudes goes beyond the scope of this paper and we leave it for future work. 

As for the possible physical mechanisms triggering the stabilization there are many possibilities. First, any perturbation of the sphaleron (or any other soliton) can be decomposed as a linear combination of the internal modes. Therefore, generic perturbations will have a nonzero superposition with the unstable mode, but also a nonzero superposition with the rest of the modes, and depending on the balance between them, one could end up with a long-living solution. In particular, the interaction of the sphaleron with radiation will lead, in general, to the internal mode excitation which ultimately could trigger the stabilization. 
It is worth mentioning that this mechanism was also found by the authors in the context of a very particular mode with false vacuum sphalerons \cite{Sergio-phi6}. This, together with the results of the present paper, suggests the universality of the phenomenon.

    
\section{Conclusions}\label{sec:conclusions}

In this paper, we have described the sphaleron decay in some relevant one-dimensional models with $\mathbb{S}^1$ as base space. First, we have reviewed the simplest sphalerons in $\phi^4$, sine-Gordon and $\phi^6$ theories. In the first two cases, the sphaleron solutions as well as their internal modes are known in an analytical form. For the $\phi^6$ theory, we found an analytical form for the sphaleron, but in this case, it seems that the internal modes cannot be obtained in a closed form. 

We have also studied the dynamics of the decay. A natural way to perform this analysis is simply to excite the sphaleron in the unstable direction, as determined by the negative mode. We have found a very rich structure of patterns during the decay. For small sphalerons in $\phi^4$, in most cases, the sphaleron forms an oscillon state. If the sphaleron is big enough, we have argued that it can be seen as a $K\bar{K}$ pair. In this situation, we have obtained a fractal pattern characteristic of the $K\bar{K}$ scattering on the real line. This shows that the resonant energy transfer mechanism, which is crucial to understanding the dynamics of solitons, plays a significant role in the dynamics of sphalerons. 

In $\phi^6$ theory, the sphaleron dynamics is even richer. As on the real line, there are inequivalent decays for the sphaleron, depending on the vacuum to which the configuration decays. On $\mathbb{S}^1$, the analog scattering between different species of kinks on the real line, can be simply accomplished by choosing the direction of excitation of the unstable mode. Again, the decay patterns of large sphalerons are very similar to the $K\bar{K}$ scattering on the real line \cite{Dorey}. It is important to note that the analogy between the decay of the sphaleron in $\phi^4$ and $\phi^6$ with the $K\bar{K}$ scattering on the real line holds at time scales of the order of the size of the base space. For longer times, the periodicity of the base space allows any mode emitted from the interaction region to interact again, transforming the configuration into a superposition of modes. This contrasts with the situation on the real line, where the possible asymptotic states resulting from the soliton scattering can be described as a superposition of kinks at certain velocities plus radiation. Here, the fact that our base space is bounded suggests that the system should have a Poincaré recurrence time. Therefore, at some possibly large but finite time, the configuration would be arbitrarily close to the initial state, leading to a sort of quasi-periodic configuration. The decay of sine-Gordon sphalerons is rather different. Due to the integrability of the model, the sphaleron decays into a time-periodic solution, which is known analytically, and, at all times, we can interpret the evolution as a periodic transition between sphalerons and antisphalerons.

We have also shown that it is possible to stabilize the $\mathbb{S}^1$ sphalerons due to the internal mode structure. The excitation of the positive internal modes leads to a positive pressure that compensates the static attraction in the direction of the unstable mode. This may increase the life of the sphaleron by producing an oscillatory solution that lasts as long as the internal mode remains excited, which, as in the standard topological soliton case, decays as $t^{-1/2}$. It should be remarked that stabilization occurs over a wide amplitude range of positive modes; therefore, it does not appear to be a fined-tuned phenomenon. We would also like to emphasize that it seems to be a rather generic mechanism that may even be triggered in more realistic scenarios, such as electroweak sphaleron, which is known to host positive internal modes \cite{Brihaye-modes}. Extensions of these results to higher dimensions are under current investigation.

\acknowledgments

J.Q. and S.N.O. has been supported in part by Spanish Ministerio de Ciencia e Innovación (MCIN) with funding from the European Union NextGenerationEU (PRTRC17.I1) and the Consejería de Educación, Junta de Castilla y León, through QCAYLE project, as well as the grant PID2023-148409NB-I00 MTM.  J.Q.  is also supported partially under the project Programa C2 from the University of Salamanca. S.N.O. acknowledge financial
support from the European Social Fund, the Operational Programme of Junta de Castilla y León and the regional Ministry of Education. Financial support of the Department of Education, Junta de Castilla y León, and FEDER Funds is gratefully acknowledged (CLU-2023-1-05).



\begin{thebibliography}{99}

\bibitem{Manton-WS} N. S. Manton, \textit{``Topology in the Weinberg-Salam theory"}, Phys. Rev. D \textbf{28}, (1983) 2019.

\bibitem{Forgacs} P. Forgács, Z. Horváth, \textit{``Topology and saddle points in field theories"}, Phys. Lett. B \textbf{138}, (1984) 397-401.

\bibitem{Manton-Klinkhamer} F. R. Klinkhamer and N. Manton, \textit{``A saddle-point solutions of the Weinberg-Salam theory"}, Phys. Rev. D \textbf{30}, (1984) 2212. 

\bibitem{Rubakov-current} V. A. Rubakov, \textit{``Unsuppressed electroweak fermion number non-conservation in decays of heavy particles"}, Nucl. Phys. B \textbf{256}, (1985) 509.

\bibitem{Rubakov-Universe} V. A. Kuzmin, V. A. Rubakov and M. E. Shaposhnikov, \textit{``On anomalous electroweak baryon-number non-conservation in the early universe"}, Phys. Lett. B \textbf{155}, (1985) 36.

\bibitem{Bochkarev-AH} A. I. Bochkarev and M. E. Shaposhnikov, \textit{``Anomalous fermion number nonconservation at high temperatures: Two-dimensional example"}, Mod. Phys. Lett. A \textbf{4}, (1989) 1495.

\bibitem{Carson-AH} L. Carson, \textit{``Approximate computation of the sphaleron prefactor. Application to the two-dimensional Abelian Higgs model"}, Phys. Rev. D \textbf{42}, (1990) 2853.

\bibitem{Brihaye-modes} Y. Brihaye and J. Kunz, \textit{``Normal modes around the $SU(2)$ sphalerons"}, Phys. Lett. B, \textbf{249} (1990) 90-96.

\bibitem{Carson-modes} L. Carson and L. McLerran, \textit{``Approximate computation of the small-fluctuation determinant around a sphaleron"}, Phys. Rev. D \textbf{41} (1990) 647.

\bibitem{Brihaye-AH} Y. Brihaye, S. Giller, 
P. Kosinski, J. Kunz, \textit{``Sphalerons and normal modes in the (1 + 1)-dimensional abelian Higgs model on the circle"}, Phys. Lett. B \textbf{293}, 3 (1992) 383-388.

\bibitem{Brihaye-AH2} S. Braibant and Y. Brihaye, \textit{Quasi-Exactly-Solvable System and Sphaleron Stability}, J.Math.Phys. \textbf{34}, (1993) 2107.

\bibitem{Yaffe-instabilities} L.G. Yaffe, \textit{``Static solutions of $SU(2)$-Higgs theory"}, Phys. Rev. D \textbf{40} (1989) 3463.

\bibitem{Akiba-instabilities} T. Akiba, H. Kikuchi and T. Yanagida, \textit{``Free energy of the sphaleron in the Weinberg-Salam model"}, Phys. Rev. D \textbf{40}, (1989) 588.

\bibitem{Sergio-phi6} S. Navarro-Obregón and J. Queiruga, \textit{``Impact of the internal modes on the sphaleron decay"}, Eur. Phys. J. C \textbf{84}, (2024) 821.

\bibitem{Bazeia-false} A. Avelar, D. Bazeia, L. Losano et al, \textit{``New lump-like structures in scalar-field models"}, Eur. Phys. J. C \textbf{55}, (2008) 133–143. 

\bibitem{Bazeia-false2} A. Avelar, D. Bazeia,  
W.B. Cardoso and L. Losano, \textit{``Lump-like structures in scalar-field models in 
 dimensions"}, Phys. Lett. A \textbf{374}, (2009) 222-227. 

\bibitem{Alberto-false} A. Alonso-Izquierdo, S. Navarro-Obregón, K. Oles, J. Queiruga and T. Romanczukiewicz, \textit{``Semi-Bogomol'nyi-Prasad-Sommerfield sphaleron and its dynamics"},  Phys. Rev. E \textbf{108}, (2023) 064208. 

\bibitem{Manton-phi3} N. S. Manton and T. Romańczukiewicz, \textit{``Simplest oscillon and its sphaleron"}, Phys. Rev. D \textbf{107}, (2023) 085012.

\bibitem{Manton-circle} N. S. Manton and T. M. Samols, \textit{``Sphalerons on a circle"}, Phys. Lett. B \textbf{207}, (1988) 179-184.

\bibitem{Jiu-modes} Jiu-Qing Liang, H. J. W. Müller-Kirsten, 
D. H. Tchrakian \textit{``Solitons, bounces and sphalerons on a circle"}, Phys. Lett. B \textbf{282}, (1992) 105-110.

\bibitem{Abramowitz} M. Abramowitz and I. A. Stegun, \textit{``Handbook of mathematical functions"}, Dover (1972).

\bibitem{Arscott-Lame} F. M. Arscott, \textit{``Periodic Differential Equations"}, Pergamon Press, New York (1964).

\bibitem{Ward-Lame} R. S. Ward, \textit{The Nahm Equations, Finite-Gap Potentials and Lamé Functions}, J.Phys. A \textbf{28},
(1987) 2679

\bibitem{Sutcliffe-Lame} P. M. Sutcliffe, \textit{Symmetric Monopoles and Finite-Gap Lamé Potentials}, J.Phys. A \textbf{29}, (1996)
5187.

\bibitem{Dorey}P. Dorey, K. Mersh, T. Romanczukiewicz and Y. Shnir, \textit{``Kink-antikink collisions in the $\phi^6$ model"}, Phys. Rev. Lett. \textbf{107} (2011) 091602. 

\bibitem{Rajamaran} R. Rajaraman, \textit{``Solitons and Instantons"}, Elsevier Science, Amsterdam (1982).

\bibitem{Sanati} M. Sanati and A. Saxena, \textit{``Half-kink lattice solution of the phi**6 model"},  J. Phys. A \textbf{32} (1999) 4311-4320.

\bibitem{Campbell} D. K. Campbell, J. F. Schonfeld, and C. A. Wingate, \textit{``Resonance structure in kink-antikink interactions in $\phi^4$ theory''}, Physica D \textbf{9}, (1983) 1.

\bibitem{Manton-energy} N. S. Manton, \textit{``An effective Lagrangian for solitons"}, Nucl. Phys. B \textbf{150}, (1979) 397.

\bibitem{costabile} G. Costabile, R. D. Parmentier, B. Savo, D. W. McLaughlin and  A. C. Scott, \textit{``Exact solutions of the sine-Gordon equation describing oscillations in a long (but finite) Josephson junction"}, Applied Physics Letters, 32(9), 587-589. 

\bibitem{marchesoni} F. Marchesoni, \textit{``Exact Solutions of the Sine-Gordon Equation with Periodic Boundary Conditions}, Progr. Theor. Phys. \textbf{77}, (1987) 813–824.

\bibitem{adam-quasi} C. Adam, K. Oles, T. Romanczukiewicz and A. Wereszczynski, \textit{``Kink-antikink collisions in a weakly interacting $\phi^4$ model"}, Phys.Rev.E \textbf{102} (2020) 6, 062214.
 
\end{thebibliography}
\end{document}